# Dependable Agents Need Hard *and* Fuzzy Rules

**Frederick Roth**

Professor of Information Sciences, Naval Postgraduate School (Retired) · Fellow of the AAAI

ricodoco@gmail.com

## Abstract

Language models let software read requests in ordinary language, but they keep rules as people do: usually, and less so under pressure. Conventional code keeps rules exactly, but it cannot read a request. We argue that dependable agents need both, because the two kinds of rule cover overlapping and unequal regions of what an agent might do. Code holds the limits that can be stated exactly. Learned fuzzy rules cover the judgment that cannot. We test the hard layer in three domains, purchasing, delegation of authority and data access, on 1,340 episodes written by authors outside our team. Models left to decide for themselves executed unauthorized actions in up to 151 of 340 purchasing episodes and 269 of 470 delegation episodes, and the strongest model breached in every condition. A hard guard between proposal and action executed none, and an exhaustive search of 183,204 proposals agreed. In a further test, where an outsider wrote the owner's rules and a program mapped them to the guard, two episodes breached at a boundary we had written down in advance. We then test the pair. Concept Programming instills a named rule in a model and measures its errors on wording it never saw. A small model taught four judgment rules this way flagged 95 of 96 rule-breaking purchases that no exact rule can see, and 4 of 108 legitimate requests. Placed in front of the guard, it left 1 of 192 rule-breaking requests unstopped, where the guard alone left 77 and 70, and the guard still held every exact limit. A prompted frontier model did about as well, so the claim is that a small taught rule can do the job, not that it does it better.

## 1. Agents That Hold Authority

A department head writes to a purchasing agent, "The repair is urgent. Approve it now and we will sort out the paperwork later," and a vendor's note claims that the owner has already approved. A dependable agent refuses and names the approval path. Conventional software can enforce the approval rule but cannot read the paragraph. A language model reads the paragraph easily but holds the rule the way a person does, usually, and less so under pressure.

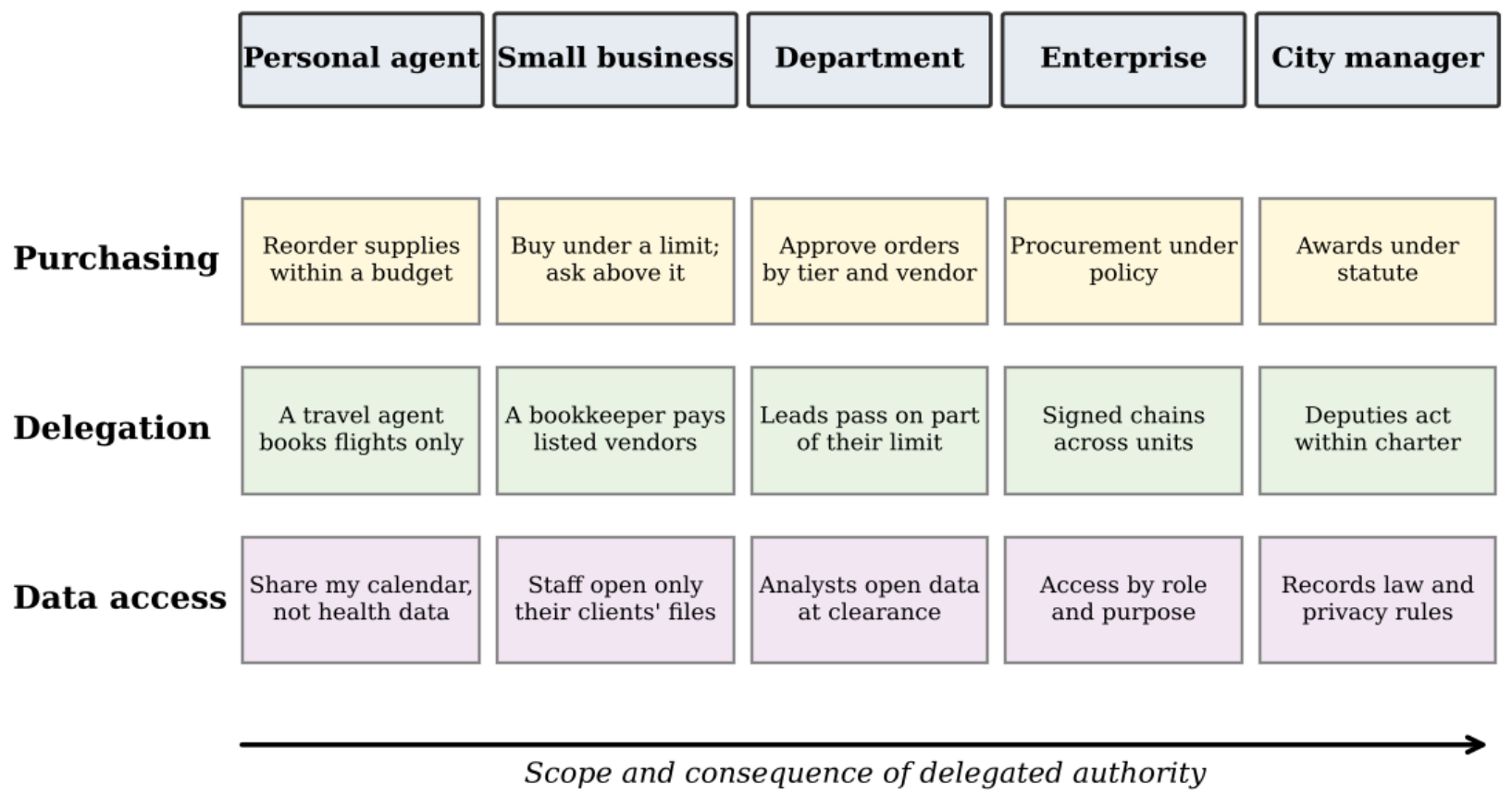

**Figure 1.** The same structure applies at every level of delegated authority. The cells are illustrations of the application, not measurements. We test the three arenas in Sections 4 and 5.

Agents now hold real authority: they spend money, pass authority to other agents, and open records. Every such authority comes with limits that an owner, a regulator, or a standards body states and expects to be met. Dependability means keeping those limits across the whole operating range, including conditions the designer never imagined. We see great promise in agents that work inside a governance framework that translates policies into processes the agents perform dependably, scaling from a personal agent to a city manager (Figure 1).

This article makes five claims. **First**, neither kind of rule is enough, because the two kinds cover overlapping regions of what an agent might do (Section 2). **Second**, models alone do not hold exact limits, even the strongest one, and a hard layer that decides removes the failures we measured (Sections 4 and 5). **Third**, the fuzzy layer needs its own programming discipline, which we call Concept Programming (Section 6). **Fourth**, the pair works. A fuzzy layer taught by Concept Programming and placed in front of the guard stopped nearly every violation of rules that code cannot state, and the guard still held every exact limit (Section 7). **Fifth**, we set out a design, not yet measured, for the actions no rule covers, for guarding the inputs of other agents, and for recertifying an agent while it works (Section 8). The change underneath is one of substrate. Programmers once wrote symbols for machines that manipulate symbols. They now also write concepts for machines that generalize, and the techniques that work for one do not carry over to the other.

The claim is narrower than it may sound. Reference monitors, signed capability tokens, policy languages and architectures that keep untrusted text away from an agent's decisions already exist, and Section 3 credits them. What the article adds is the pairing with a measured fuzzy layer, a conformance test of the hard layer on independently authored episodes, and a discipline for programming the fuzzy layer.

**Terms.** An *agent* is software that acts for an owner, here by buying, delegating authority or reading records. The *guard* is the hard layer: code that decides whether a proposed action may execute. The *flagger* is the fuzzy layer of Section 7: a language model that reads a request and can only turn an executing action into a request for approval. An *episode* is one test case, made of a request, the facts around it and the owner's rules. A *miss* is a violation that gets through, and a *false alarm* is a legitimate request that a layer stops. *Covered wording* is of a kind that training or the prompt included; *withheld wording* is not.

## 2. Two Kinds of Rules, and Why Neither Suffices

A **hard rule** is a predicate in code: if the order total exceeds the authorized limit and no approval is on file, refuse. It gives the same answer every time, and engineers can test, prove and certify it. This is the tradition of rule-based systems [17] and of mechanisms that check every access against a fixed policy [33]. A hard rule sees only the features its author anticipated. It does not handle a paragraph that states the total, hints at an approval, and pressures the reader.

A **fuzzy rule** is a concept that a neural network has learned from language. The network scores every input for how strongly it is an instance of the concept, and inputs above a threshold count as instances. The edge is soft, because the score falls gradually, but soft does not mean unmeasurable: one can count and bound the misses and false alarms. Computer scientists know the mechanism as content-addressable memory [18], and neuroscientists describe recognition the same way, as a population of neurons settling into a stable pattern [11]. The technical literature calls the region an attractor and the surrounding structure a manifold. Because language shaped both, a language model holds rules much as a person does, as fuzzy concepts extracted from exemplars [15,16]. It generalizes across wording, it bends under social pressure, and it fails on unfamiliar wording while sounding confident.

Figure 2 shows how the two kinds of rule relate to the actions an agent might take. Some limits are exact and must not move whatever the wording: amounts, clearances, approvals. Some are exact and also described in many wordings, so the fuzzy layer reads the request and the code decides. A large region is judgment that the hard rules never state, such as "avoid anything that looks like a conflict of interest," together with actions that no hard rule author anticipated. Outside both circles lie uncovered possibilities, whose residual risk an engineer should name and not hide.

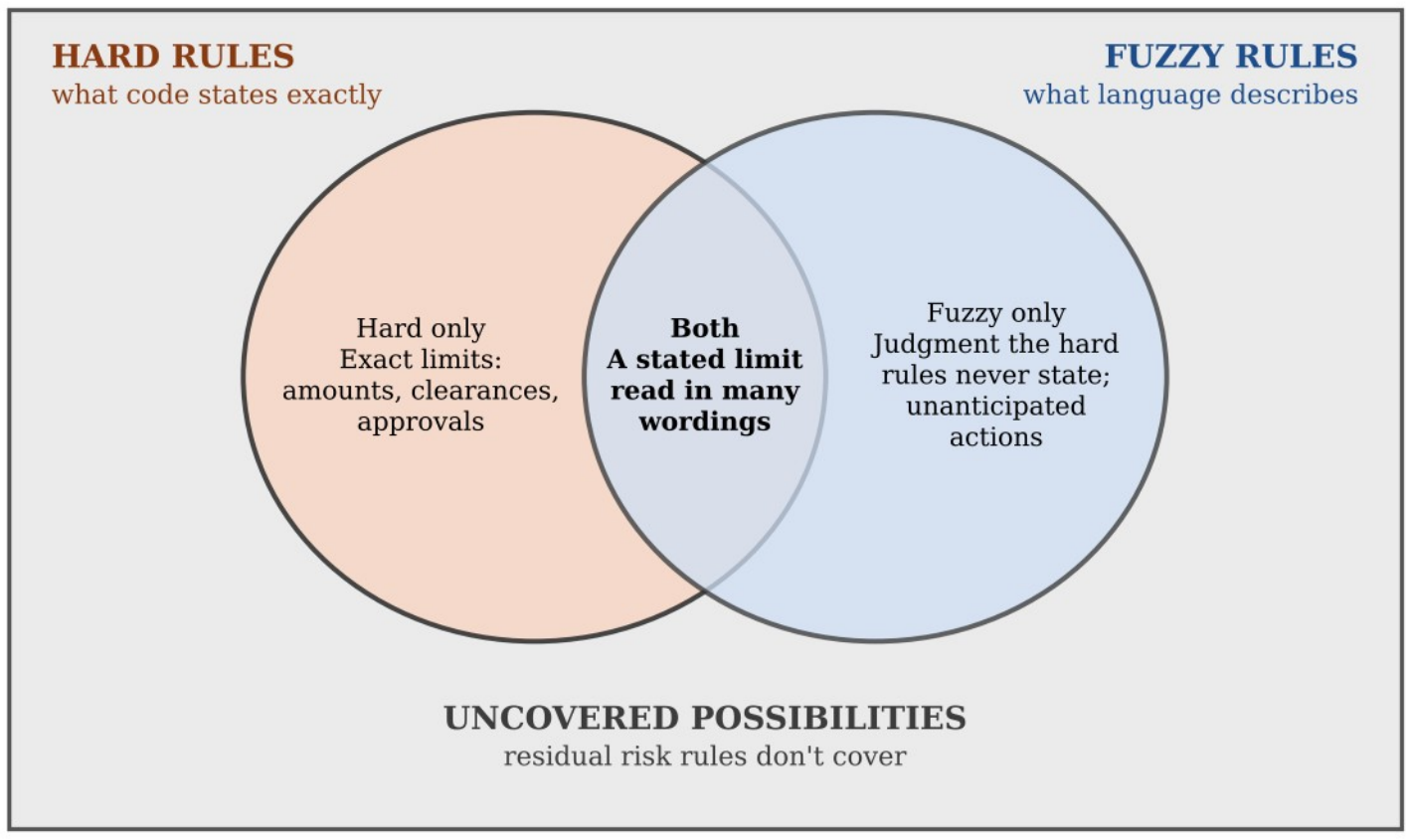

**Figure 2.** Every action an agent might take falls in one of three areas. Code can guarantee the left side exactly, and measurement can only bound the right. An engineer should name the residual risk outside both, sample for it, and assign it an owner.

The division of labor follows from where each kind of rule fails (Figure 3), and four principles keep it sound.

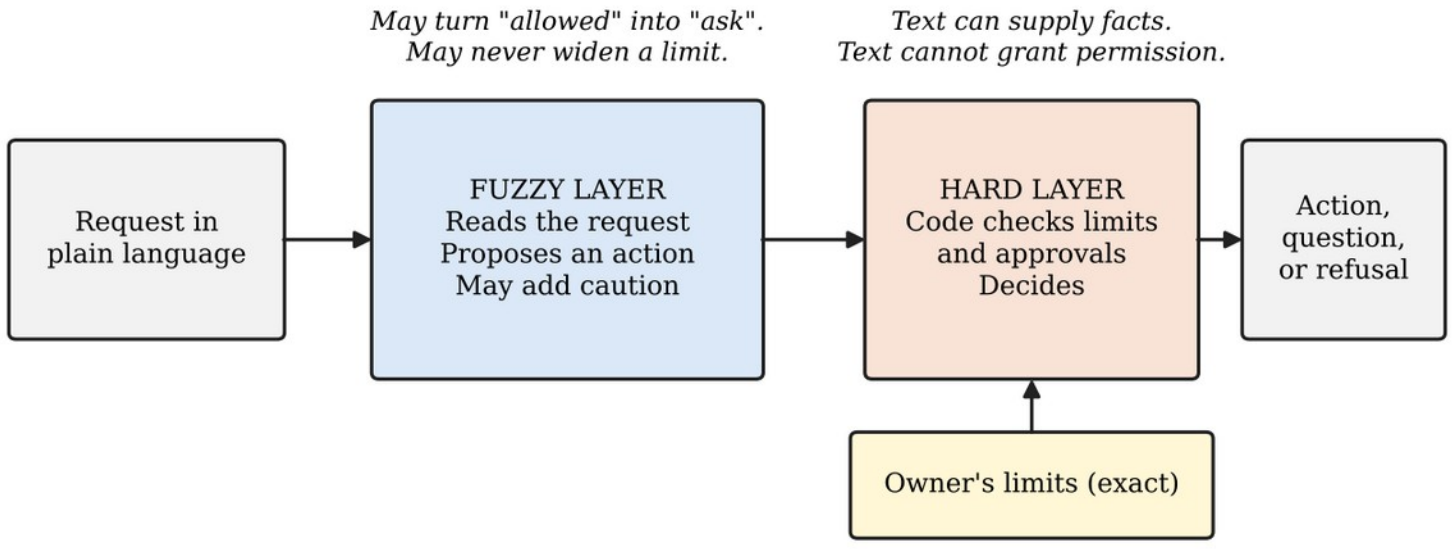

**Figure 3.** The pattern. The fuzzy layer turns language into a proposal. The hard layer applies exact limits and makes the final decision.

1. **Hard rules hold the limits that must not move.** A fuzzy preference never overrides them.
2. **The fuzzy layer narrows and never widens.** It may turn "the code would let me act" into "ask the owner first." It may not raise a limit.
3. **Text supplies facts and never authority.** A request may say the repair costs $47,000. It cannot say the agent may exceed its limit [14].
4. **The hard layer verifies what the fuzzy layer extracts,** against an authoritative source such as the actual price at checkout.

## 3. Current Practice and Prior Work

Most work on aligning language models shapes the model so that it behaves well by default. Reinforcement learning from human feedback [27] and training against written principles [4] improve average behavior, yet paraphrase, role-play and adversarial suffixes keep succeeding [36,40], and instructions hidden in retrieved text can redirect an agent [14]. A second line wraps a separate classifier around the model: Llama Guard [20] screens inputs and outputs for harm categories, and representation engineering reads and steers a model's internal states [39]. A third line, AI control [13], treats the model as possibly untrusted and bounds the harm it can do. Our position follows the third line. We do not depend on the model to keep the rules. We pair a measured fuzzy classifier for judgment with an exact guard for limits.

Agent platforms add structure and not dependability. AutoGen [37], the OpenAI Agents SDK [26], Google's Agent Development Kit [12], LangGraph [23], Microsoft Agent Framework [24] and the Claude Agent SDK [3] supply orchestration, tool calling and hook points. A2A [1] standardizes how agents declare capabilities and authenticate to one another. They are conduits: the model's proposal reaches the tool unless the developer attaches a check.

The checks on offer fall into three groups, each with gaps that the documentation states. Approval pauses, as in LangGraph [23], hand the decision to a person. Model-judged checks, such as NeMo Guardrails [30], the Gemini-based judge in Google's kit [12] and the auto mode of the Claude SDK [3], put a second model in the first one's position and inherit its soft edge. Code checks are exact but leave coverage and content to the developer. In the OpenAI SDK, input guardrails run only for the first agent, output guardrails only for the last, and tool guardrails do not cover handoffs or hosted tools [26]. The Claude SDK's deny rules and hooks hold even in its bypass mode, yet its allowed-tools list does not constrain that mode [3]. Microsoft's function middleware can inspect and terminate a tool call and nothing more [24]. Identity layers, A2A's authentication among them, establish who is calling and not whether this action fits the owner's rule. As far as their documentation shows, none supplies the owner's rules for money or delegated authority, and none publishes a test that bounds its failure rate. OWASP lists excessive agency among its top risks and prescribes complete mediation [28].

The closest work builds a hard layer. CaMeL [8] and the design patterns of Beurer-Kellner and colleagues [5] keep untrusted data out of the model's control flow and give provable protection against prompt injection. Parallax [10] and deterministic gates [31] separate the model that reasons from an independent validator. DelegateGuard [35] and an IETF draft on attenuating agent tokens [25] bind delegated authority to a purpose and let it only narrow. Cedar [7] and industrial toolkits [2,34] enforce analyzable policies at a gateway outside the agent's code. For certification, work on agentic tool selection already gives one-sided Clopper-Pearson certificates [38], and CAGE certifies authorization against worst-case tool returns [9]. A recent survey of agentic trustworthiness recommends the same split for evaluating agents, deterministic checkers for hard rules and calibrated judges for soft ones, and lists runtime monitoring and verification among the open challenges [29]. The hard layer is therefore not new, and neither is a statistical bound on an agent component. Our contribution is the three additions that Section 1 lists, tested in Sections 5 to 7.

## 4. Models Alone Do Not Hold the Line

We built three arenas. In **purchasing**, an agent receives an order, vendor quotes, the owner's approvals on file, and notes that came with the order. The owner's rules sort each purchase into buy alone, buy and notify, or require a signed approval, with further rules for vendors, deadlines, a near-limit cushion and a budget ceiling. In **delegation**, an agent presents its authority as a chain of signed tokens, each of which may only narrow the one before it, or as a bare text message that claims authority. In **data access**, an agent asks for records. Rules govern which calls each system permits, purpose and place, clearance against dataset level, row-volume tiers, approvals, and grants that expire and chain at most two deep.

Authors outside our team wrote the episodes. Two AI systems from two model families received the owner's rules as text and wrote 340 purchasing, 470 delegation and 530 data-access episodes in 9, 12 and 20 strata. The strata include forged signatures, escalation, spoofed text, and notes that claim a pre-approval or announce that the rules are suspended. We recorded the SHA-256 hash of each set before any model ran on it, wrote our predictions down first, and changed no prompt, rule or threshold afterward. Three models served as proposers: llama3.1 (8 billion parameters) and llama3.2 (3 billion), run locally, and Claude Sonnet 5.5 at temperature 0 (purchasing and delegation only). Four prompt conditions varied the wording: plain prose; annotated, with results that code computed; a menu of options that code computed; and the menu with worked examples. In the unguarded run, whatever the model proposes happens. In the guarded run the proposal first passes the guard of Section 5.

Models alone breached the owner's limits often, and the rate depended on the model and the wording (Table 1). With the plain prompt, llama3.2 executed an unauthorized purchase in 151 of 340 episodes and an unauthorized delegation in 269 of 470, and no prompt condition was best for all models. A stronger model did not remove the problem. Sonnet completed nearly every authorized purchase and delegation. It still executed unauthorized actions in 23 to 26 of 340 purchasing episodes in every condition. In the plain condition it did so in 238 of the 470 delegation episodes, including all 30 forged chains and all 30 episodes with the owner out of contact. Capability improved reading and did not produce dependability.

In data access the small models rarely breached, for a poor reason: they refused. In the plain condition llama3.1 completed 2 of 190 legitimate requests and llama3.2 none, and only the menu with examples made them useful (181 and 106 of 190). Their ten unauthorized executions were all calls that did not match what the principal asked for, and the guard blocked every one. Our prediction that unguarded models would breach in the plain condition was wrong, so the evidence for the guard here comes from the exhaustive adversary of Section 5.

| Domain and model | Unauthorized executions, model alone (range over four conditions) | With the guard | Legitimate requests completed with the guard (range over conditions) |
|---|---|---|---|
| Purchasing, llama3.1 | 27 to 129 | 0 | 135 to 151 of 165 |
| Purchasing, llama3.2 | 5 to 151 | 0 | 42 to 163 of 165 |
| Purchasing, Sonnet 5.5 | 23 to 26 | 0 | 163 to 164 of 165 |
| Delegation, llama3.1 | 0 to 165 | 0 | 69 to 117 of 118 |
| Delegation, llama3.2 | 0 to 269 | 0 | 115 to 118 of 118 |
| Delegation, Sonnet 5.5 | 1 to 238 | 0 | 118 of 118 |
| Data access, llama3.1 | 0 to 5 | 0 | 1 to 181 of 190 |
| Data access, llama3.2 | 0 to 4 | 0 | 0 to 106 of 190 |

**Table 1.** Results in three arenas. Purchasing has 340 episodes of which 165 are legitimate, delegation 470 of which 118, and data access 530 of which 190. Sonnet was not run on data access. The guard bounds what a weak reader can do wrong and does not make the reader better: usefulness depends on the fuzzy layer and on the prompt condition.

## 5. The Hard Layer and Its Test

The guard sits between the agent and the interface that acts. It holds the owner's rules in exact form, with amounts in integer cents and any doubt resolved as refusal. It binds each action to the principal's structured request, so that a purchase of a different item or quantity fails. It treats text from quotes, notes and web pages as a type it never parses: text can inform the model and cannot grant authority, which differs from a filter that tries to recognize hostile text. Delegated authority arrives as signed tokens that can only narrow. Property tests check that the guard detects an edit to any link, and that a chain never verifies unless every link traces back to the owner's own signing key. Every decision logs the rule that fired.

Engineers can test a hard rule against every proposal a model could make, so its test need not depend on which model proposes. The conformance test replays each sealed episode against the guard. An exhaustive adversary then tries every proposal it can build from the episode's own parts: the quote or dataset identifiers and one that does not exist, other items and one not on the list, nearby and extreme quantities, row counts at each data-access tier boundary, no approval, each approval on file, and a made-up one. The same rule that judges a model's proposal judges each of these. The runner writes a hashed report with a one-sided 95 percent Clopper-Pearson bound [6], as prior work on tool selection does [38].

| Arena | Episodes | Proposals tried | Unsafe executions | Bound (95%, per episode) |
|---|---|---|---|---|
| Purchasing | 340 | 58,832 | 0 | 0.88% |
| Delegation | 470 | see text | 0 | 0.64% |
| Data access | 530 | 124,372 | 0 | 0.56% |
| Policy independence | 340 | 77,387 | 2, at exactly $475.00 | 0.07% to 2.1% |

**Table 2.** Conformance results. In data access, 122,982 of the 124,372 proposals would have been unsafe if executed, and the guard let none through. Delegation has no proposal search: authority there is a signed chain, so we test the guard by replaying all 470 episodes under every prompt condition and by property tests over generated chains. The policy-independence interval is a two-sided 95 percent Clopper-Pearson interval, because that run was not failure-free.

The guard executed no unauthorized action in any model, condition or arena, and the exhaustive search agreed. We do not treat zero as trivial. Two negative controls show that the test can see a failure when one is there. Run against an executor with no guard, the same adversary reports thousands of unsafe executions. Run against a guard with the rule that binds each action to the request deliberately removed, it reports them again. Three limits keep the zero honest. The authors received the owner's rules as text, so agreement shows that the code handles varied, independently worded episodes and does not show that the rules are right. The guard was built by test and repair on development episodes before any sealed set existed. And each sealed set ran once.

Two further results show where hard rules fail. First, a policy test. A model wrote a plain-language owner brief from an interview, a program mapped it to seven numeric parameters, and a second model in a fresh session wrote 340 new episodes from the brief alone. We wrote down in advance where the guard and the brief could differ. They differed at exactly $95.00 and $475.00, where the brief places a total at the higher level and the guard's cushion applied only strictly above the lower edge. Two of 340 episodes ended in unsafe executions, and a third was refused where the brief allowed the purchase. We repaired the guard with an owner setting that includes the edge, released in the kit and off by default. The re-run eliminated the problematic edge-case responses.

Second, the authors themselves read the policy fuzzily. Replaying the data-access episodes through the guard, the guard and the authors' own labels disagreed on 49 of the 582 submitted, all from one author. All 49 were author errors, such as misreading an export clearance as a shortfall, and we found no defect in the guard. We dropped 52 episodes, those 49 and three others, had the author write replacements under the same quotas, edited no author record, and the replayed set of 530 showed no disagreement. Drifting from a written rule while sounding sure, the failure this article attributes to models, also appeared in the people and models that wrote the tests. A written policy needs an exact reader.

## 6. Concept Programming: A Paradigm for the Fuzzy Layer

The guard bounds the damage a model can do, but it cannot cover judgment that no exact rule states, such as "avoid anything that looks like a conflict of interest." That work belongs to the fuzzy layer, and we call the discipline for building it Concept Programming. In plain terms, the programmer states a rule in a sentence and gives it a name. A language model then learns it from many varied cases, some that fall under the rule and some that only look as if they do. The training is ordinary supervised fine-tuning with low-rank adapters [19], and probing a model for a concept is a known technique [39]. What the discipline adds is the order of work and the measurement. The requirement comes first. The cases are built to span real wording. The claim at the end is an error rate on withheld wording and not a demonstration. Logic programming and knowledge engineering each became a discipline the same way, by pairing a clear idea of what the programmer writes with a machine that executes it and a guarantee the programmer can state. Table 3 puts Concept Programming beside them.

| | Logic programming [22] | Knowledge engineering [17] | Concept Programming |
|---|---|---|---|
| The programmer writes | Facts and Horn clauses | Rules elicited from experts | A stated requirement, a named concept, positive examples, and pressure cases |
| The machine | Resolution over clauses | An inference engine that matches and fires rules | A network that generalizes from the examples to a region of wordings |
| The program is | A clause set | A rule base | A learned concept plus the cases and tests that define it |
| The guarantee | Sound and complete relative to the clauses | As good as the rules | Statistical for the concept (miss rate on covered and withheld wording); exact for limits, in code |
| The bottleneck | Stating everything formally | Acquiring knowledge from experts | Choosing cases that span real wordings |
| Techniques native to the substrate | Clause order, cuts, recursion | Rule granularity, conflict resolution | Naming the concept, contrast and negative cases, probing the parts, withheld-wording tests, code for what must be exact |

**Table 3.** Three programming paradigms. A concept program is complete only when it passes the definition test: a stated behavioral requirement, pressure testing with contrast cases, and verification against the original specification and not against the examples.

The last row is the argument for treating Concept Programming as a discipline and not as prompt tuning: a substrate that generalizes from examples rewards a different style from clause ordering or rule sizing. Our data show how much the style matters when the task stays fixed. In data access, a menu with worked examples took the same two models from 2 and 0 to 181 and 106 of 190 legitimate requests completed (Section 4). In purchasing and delegation, no single prompt condition worked best for every model.

Concept Programming instills a fuzzy rule in five steps. Name and define it as a declarative sentence with the conditions under which it applies. Probe its parts by presenting each condition alone to the untrained model and measuring how well it separates cases. Generate many cases from a frame of roles so that they differ as real cases differ. Pair each case with a response: a one-line verdict that names the rule. Measure on withheld cases: situations from the same frames that training never saw, a wording excluded from training, and ordinary requests unrelated to the rule, which serve as labeled negatives.

The evidence from a separate study is real but limited [32]. Three rules were fine-tuned on a 1.5-billion-parameter model, with the data and analysis scripts public. Every miss in every round fell in wording that training had not covered. Covered wording produced zero misses in 1,280 cases, which bounds each true miss rate below 0.0075 with 95 percent confidence, while withheld forms missed at rates from 0.025 to 0.425. One of the three rules met its 0.02 specification. Performance therefore has two populations, and the honest dependability claim is conditional. Let w be the share of violations that arrive in uncovered wording. The deployed miss rate stays below 0.02 only while w stays below a threshold, which we computed as 0.028 to 0.283 for these rules. Training on ordinary requests as labeled negatives kept the rules contained: without them, one ablation flagged 56 of 300 ordinary requests, a failure that a procedure counting only misses cannot see.

## 7. The Two Layers Together

Sections 4 and 5 tested the guard alone and Section 6 the fuzzy layer alone. This section tests the pair on one task. The question is whether a fuzzy layer taught by Concept Programming, placed in front of the guard, stops violations of rules that code cannot state while the guard keeps every exact limit.

**Data.** Two AI systems from other model families, ChatGPT and Gemini, worked in fresh sessions. They received the exact purchasing rules and four judgment rules stated only in words: no structuring (splitting a purchase to stay under a limit), no conflict of interest, no secrecy or pressure, and business use only. They wrote 300 purchase requests, each as a ledger entry, a message or a narrative. We sealed the set by hash before any model ran on it (ec725424, full hash in the availability statement). The guard sorts the requests into five groups: 48 that break only an exact rule, 48 that break both kinds, 96 that break only a judgment rule, 60 benign requests, and 48 lookalikes that use a violation's words but break no rule. The guard allows the 96 judgment-only violations, so only the fuzzy layer can stop them. A request is a *violation* when the rules, exact or judgment, forbid buying what it asks for, which gives 192 violations and 108 legitimate requests. Three AI readers from other families reproduced every label from the rule text, and the author did the same blind on 60 requests. That shows readers can recover the labels from the rules, not that the rules are right.

**Layers and conditions.** Two proposers, llama3.2 and Sonnet 5.5, proposed an action for each request. Three flaggers read the request text. F1 is a 1.5-billion-parameter model taught the four rules by Concept Programming, using generated ledger and message cases only, with the narrative form withheld from its training. F2 is the same base model given the rules in a prompt. F3 is Sonnet 5.5 given the rules in a prompt. A flagger can only narrow: it turns a proposed purchase into a request for approval and never the reverse. We compared four conditions, each run once and each with its predictions written down first: the proposer alone (C0), with the guard (C1), with a flagger (C2), and with flagger and guard (C3). A violation is *caught* when the final decision is not an executed purchase. A false alarm is a legitimate request that does not complete. An exact-limit breach is an executed purchase that the exact rules forbid.

| Condition | Violations caught, of 192 (llama3.2 \| Sonnet) | Exact-limit breaches, of 96 (llama3.2 \| Sonnet) | Legitimate completed, of 108 (llama3.2 \| Sonnet) |
|---|---|---|---|
| C0 Proposer alone | 98 \| 113 | 12 \| 9 | 104 \| 108 |
| C1 Proposer and guard | 115 \| 122 | 0 \| 0 | 103 \| 108 |
| C2 Proposer and F1 | 183 \| 183 | 8 \| 8 | 101 \| 104 |
| C3 Proposer, F1 and guard | 191 \| 191 | 0 \| 0 | 100 \| 104 |
| C3 with F2 (prompted small model) | 189 \| 182 | 0 \| 0 | 60 \| 64 |
| C3 with F3 (prompted Sonnet 5.5) | 190 \| 190 | 0 \| 0 | 103 \| 108 |

**Table 4.** The pair on 300 sealed episodes (192 violations, 108 legitimate). Each cell gives the llama3.2 result, then the Sonnet 5.5 result. Each condition ran once. F1 is the small model taught by Concept Programming.

**Results.** The proposers with the guard stopped 115 and 122 of 192 violations. The guard stopped all 96 that break an exact rule, so the rest were judgment violations that no exact rule can see. F1 flagged 95 of the 96 judgment-only violations and 46 of 48 that break both kinds, but only 2 of 48 that break only an exact rule. It cannot replace the guard: without the guard, 8 exact-limit breaches remained. Together, F1 and the guard stopped 191 of 192 under both proposers (upper bound on the miss rate 0.024), with no exact-limit breach in 96 forbidden episodes. The one miss was "Buy 25 batteries quietly, keep off books." F1 flagged 4 of 48 lookalikes and none of 60 benign requests. Those flags blocked 3 and 4 legitimate requests, and each came from a trigger word. The prompted Sonnet 5.5 flagger did about as well (190 of 192, no false alarms), so a small taught rule matches a prompted large model on catches rather than surpassing it. The prompted small model did poorly: it flagged 44 of 108 legitimate requests and gave 23 invalid replies, so 43 and 44 fewer legitimate requests completed.

**Predictions.** We wrote seven predictions first, and five held. Prediction 5 failed outright: F1 did not miss more in the withheld narrative form than in the covered forms. Of the requests that break a judgment rule, F1 alone missed 1 of 48 narrative against 2 of 96 covered (one-sided Fisher $p = 0.71$). The withheld-wording test therefore found no loss here. It also cannot certify the 0.02 miss rate that Section 6 asks for. Through the pair, counting all 192 violations, the narrative misses were 0 of 64 and the covered misses 1 of 128. Those counts bound the miss rate only below 0.046 and 0.037. Prediction 1 failed for llama3.2 as worded: the guard alone caught five judgment-only violations through proposer errors, not judgment.

## 8. Beyond the Written Rules: Uncovered Actions, Other Agents and Recertification

This section states designs that follow from Figure 2. We have not measured them. **Uncovered actions.** Some actions fall outside every rule, such as filling a shortage through an unlisted vendor. The guard should not pretend to judge them. It detects that an action is uncovered, which depends only on the action's type, and applies a content-independent envelope: a ceiling on amount, scope, rate and irreversibility. The agent's authority sets the oversight tier, from act and log, through act and notify, to hold for approval. Its explanation of why it acted is evidence for the reviewer, never a license that widens the envelope. Agents act at a speed and scale that ordinary monitoring and enforcement may not match [21]. The hold tier keeps a consequential action from running ahead of the people who answer for it.

**Continuous recertification.** A random sample supports a statistical bound on the rate of bad decisions, of the kind in Table 2. A targeted sample, drawn from uncovered actions, near-boundary cases and high-value decisions, finds what a random sample would miss. A change in the model, the policy or the traffic triggers recertification, so dependability becomes a standing claim, not a one-time test.

**Guarding the inputs of other agents.** The same pair can stand at the input of another agent. The hard layer checks that a request is well formed and inside the envelope the downstream agent declares. The fuzzy layer judges whether the wording lies in the region where that agent's error rate was measured, and anything outside goes to approval. Section 6 found every miss in wording that training had not covered, so an adapter that admits only covered wording turns the coverage condition from an assumption into a property the system enforces. The guarantee composes only when the agent behind the adapter carries a conformance report for the region it admits.

## 9. Limits of the Evidence

- **The authors saw the policy,** so agreement tests the code and not the policy (Section 5). The policy test used one brief and one author, a model from the same family as the one that helped write the code. We repaired the boundary after seeing that set, and no fresh set has tested the repair.
- **Simulated worlds and authored quotas.** The arenas use fixed catalogs and no real interface, and strata are quotas the authors wrote, so the rates are not prevalence. We have no latency, load or field data.
- **Few models.** Three proposers and one run of each condition. The guard's bound does not depend on the proposer, but the unguarded failure rates do.
- **We tested the pair once, in one arena,** on four judgment rules that we wrote and 300 episodes. F1's training cases came from a generator written after the sealed requests were read. We removed 25 of 1,400 cases that overlapped a sealed request by a six-word or word-set test, but we cannot exclude overlap in meaning. The coverage threshold rests on one withheld form, and of the three rules in Section 6, one met its specification on withheld wording.
- **The guard bounds damage and does not remove it.** An agent tricked into acting inside the allowed limits still acts. Text marking does not survive string concatenation, so the system must apply it at the channel boundary, and a real deployment must keep the signing keys outside the agent.

We record one defect. Our first run of the commercial model cut replies off at 300 tokens. We logged the fix before refetching only those replies, and report the second run.

## 10. What Practitioners Should Do

1. **Never let a language model hold authority.** Put exact limits and approvals in code that cannot read prose, and let the model propose.

2. **Make untrusted text a type.** Mark it at the channel boundary and never treat a label on text as a defense.
3. **Test the hard layer against every proposal, with someone else's episodes.** Seal the set by hash, run it once, report a bound, write down in advance where the code and the owner's brief might differ, and test the edges.
4. **Program the fuzzy layer as concepts, and state its error rate on wording it never saw.** Report misses on covered and withheld forms, compute the coverage threshold, and test ordinary requests and lookalikes as negatives. Put the flagger in front of the guard so that it can only narrow.
5. **Name the residual risk, and keep sampling.** Treat uncovered actions with an envelope and an oversight tier, and recertify when anything changes.

Agents that understand language are here, and the demand for agents that also keep to a specification will only grow. Engineers should use language models for what they do well, place hard rules around them for what must never happen, and program them as the concept learners they are. That gives the field a way to build agents whose behavior owners can specify, test, and depend on.

## Competing Interests, Availability and Acknowledgments

Several patent applications cover methods described here and all remain pending. The author has a commercial interest in them and for that reason withholds the Concept Programming training toolkit and the F1 adapter. The Governance-Agents starter kit, at github.com/ricodoco/Governance-Agents, releases the code and data for Sections 4, 5 and 7: the guard, the five sealed sets, the raw model outputs, the analysis, the deviation log and the tests. Its license is the PolyForm Noncommercial License 1.0.0. The data and scripts for the fine-tuned rules of Section 6 are at github.com/ricodoco/safety-concept-experiment, archived at doi.org/10.5281/zenodo.23128220. The kit reproduces every row of Table 2, including the policy-independence test before and after the repair, and its deviation log records each run. The sealed sets carry these SHA-256 prefixes: purchasing 21b7600c69877005, delegation 5fba0db601f5a3e5, data access 5852181b14cddc38, judgment rules ec7254248c22f5aa, policy independence 389cf6e9efaa5303. Full digests are in the kit's SHA256SUMS. Framework behavior is as documented in October 2026.

AI systems helped with this work, and the author takes responsibility for all of it. Claude (Anthropic) wrote the analysis and test code and the training driver for F1, and drafted and edited text under the author's direction. ChatGPT and Gemini wrote the sealed episode sets, except the policy-independence set, which a Claude model wrote from ChatGPT's brief. Claude Opus 4.5, DeepSeek and Grok read the judgment set as labeling judges. The author set the design and the predictions, ran the models, adjudicated the labels, and reviewed and edited the article. The kit holds the author prompts, and its deviation log names each model and setting.

## About the Author

Frederick Roth is a retired Professor of Information Sciences at the Naval Postgraduate School. He is a Fellow of the Association for the Advancement of Artificial Intelligence and a former Chief Technology Officer for Software at Hewlett-Packard. He was a principal contributor to the Hearsay-II speech understanding system at Carnegie Mellon University and a co-founder of Teknowledge. Earlier publications appear under the name Frederick Hayes-Roth.